\documentclass[twocolumn,showpacs,preprintnumbers,amsmath,amssymb]{revtex4}

\usepackage{graphicx}
\usepackage{dcolumn}
\usepackage{bm}
\usepackage[dvips,usenames,dvipsnames]{color}
\usepackage[latin1]{inputenc}

\begin{document}
\title{Universal Quantum Computation Using Continuous Dynamical Decoupling}
\author{F. F. Fanchini}
 \email{fanchini@ifi.unicamp.br}
\affiliation{Instituto de F\'{\i}sica Gleb Wataghin, Universidade
Estadual  de Campinas, P.O. Box 6165, CEP 13083-970, Campinas, SP,
Brazil}
\author{R. d. J. Napolitano}
\affiliation{Instituto de F\'{\i}sica de S\~{a}o Carlos,
Universidade de S\~{a}o Paulo, P.O. Box 369, 13560-970, S\~{a}o
Carlos, SP, Brazil}
\author{A. O. Caldeira}
\affiliation{Instituto de F\'{\i}sica Gleb Wataghin,  Universidade
Estadual de Campinas, P.O. Box 6165, CEP 13083-970, Campinas, SP,
Brazil}
\date{\today}

\begin{abstract}
We show, for the first time, that continuous dynamical decoupling can preserve the
coherence  of a two-qubit state as it evolves during a $\sqrt{\rm
SWAP}$ quantum operation. Hence, because the Heisenberg exchange
interaction alone can be used for achieving universal quantum
computation, its combination with continuous dynamical decoupling
can also make the computation robust against general environmental
perturbations. Furthermore, since the exchange-interaction
Hamiltonian is invariant under rotations, the same control-field
arrangement used to protect a stationary quantum-memory state can
also preserve the coherence of the driven qubits. The simplicity
of the required control fields greatly improves prospects for an
experimental realization.
\end{abstract}

\maketitle

Quantum computers use superposition and entanglement of qubits  to
outperform digital computers \cite{deutsch92}. The advent of these
machines will unquestionably encompass a radical transformation in
the way we simulate quantum-mechanical processes \cite{feynman82},
imparting a plethora of new achievements in science and
technology. However, the benefits of reliable quantum information
processing depend on the development of efficient ways to avoid or
recover from qubit errors induced by environmental interaction
\cite{zurek03}.

A universal set of quantum gates consists of arbitrary
single-qubit  coherent rotations and a particular entangling
operation \cite{divincenzo95}. Such a two-qubit unitary operation
is the $\sqrt{\rm SWAP}$ gate, that has recently been realized
experimentally using double quantum dots \cite{petta05} and
neutral atoms in an optical lattice \cite{anderlini07}. An ideal
$\sqrt{\rm SWAP}$ gate is obtained by the Heisenberg coupling
between two qubits, whose dynamics is governed by the Hamiltonian
\begin{eqnarray}
H_{0}=J{\bm \sigma}^{(1)}\cdot{\bm \sigma}^{(2)},\label{H0}
\end{eqnarray}
where we use $\hbar =1$ throughout, $J$ is the exchange constant,
and, for $s=1,2$, ${\bm \sigma}^{(s)}={\bf
\hat{x}}\sigma_{x}^{(s)}+{\bf \hat{y}}\sigma_{y}^{(s)}+{\bf
\hat{z}}\sigma_{z}^{(s)}$, where $\sigma_{x}^{(s)}$,
$\sigma_{y}^{(s)}$, and $\sigma_{z}^{(s)}$ are the Pauli matrices
acting on qubit $s$. Remarkably, it has been shown that the
Heisenberg interaction alone is enough for universal quantum
computation, without the need of supplementary single-qubit
operations \cite{divincenzo00}. Thus, a protective scheme for
quantum gates such as the $\sqrt{\rm SWAP}$ is of fundamental
importance, since any quantum computation can be based solely on
the exchange interaction.

Here we show, for the first time, the effectiveness of dynamical decoupling to protect the
$\sqrt{\rm SWAP}$ quantum-gate operation on physical qubits, during a dynamical evolution subject to a noisy environment. Our procedure admits, not only pulsed,
but also continuous control Hamiltonians and, in fact, can culminate in a better protection \cite{clausen}.
Since the Heisenberg interaction is given by a scalar product, it  is
invariant under rotations. The implication of such a rotation invariance is that a simple field arrangement suffices to protect, simultaneously, not only the $\sqrt{\rm SWAP}$ quantum-gate operation, but also a quantum memory. Furthermore, the protection thus achieved is against general classes of errors, granting increased chances of successful experiments.

%Accordingly, in the present work,
%instead of treating the protection or correction of the effective
%quantum operations on the logical qubits,
%we show the effectiveness of continuous dynamical decoupling to protect the
%$\sqrt{\rm SWAP}$ operation on the physical qubits. It is
%important to emphasize that our procedure admits, not only pulsed,
%but also continuous control Hamiltonians and that, in fact,
%continuous driving can culminate in a better protection
%\cite{clausen}.

In a previous work, we have shown that it is possible to use a
continuously-applied external field to protect entangled states
from errors caused by the unavoidable interactions between the
qubit system and its environment \cite{fanchini07}. The question
naturally arises as to whether it is also possible to protect an
entangling operation. Here we show that the very same
external-field configuration of Ref. \cite{fanchini07} can prevent
errors from occurring during the application of a $\sqrt{\rm
SWAP}$ quantum gate. In other words, we show that if the control
Hamiltonian is written as
\begin{eqnarray}
H_{c}(t)={\bm \Omega}(t)\cdot\left( {\bm \sigma}^{(1)}+{\bm \sigma}^{(2)}\right),
\end{eqnarray}
then the dynamics is protected by the field arrangement given by
\begin{eqnarray}
{\bm \Omega}(t)={\bf \hat{x}}n_{x}\omega +n_{z}\omega \left[{\bf
\hat{z}}  \cos\left( n_{x}\omega t \right)-{\bf \hat{y}}
\sin\left( n_{x}\omega t \right) \right] ,\label{Omega}
\end{eqnarray}
where $\omega =2\pi/t_{c}$, $n_{x}$ and $n_{z}\ne n_{x}$ are
non-zero  integers, and $t_{c}$ is a constant. This is a simple
combination of a static field along the $x$ axis and a rotating
field in the $yz$ plane. Moreover, addressing each qubit
independently is not necessary; the field is supposed to be
spatially uniform in the neighborhood surrounding both qubits.

The evolution operator associated with the control Hamiltonian is as in Ref. \cite{fanchini07}:
\begin{eqnarray}
U_{c}(t)=U^{(2)}(t)U^{(1)}(t)=U^{(1)}(t)U^{(2)}(t),\label{Uc}
\end{eqnarray}
since ${\bm \sigma}^{(1)}$ and ${\bm \sigma}^{(2)}$ commute, where
\begin{eqnarray}
U^{(s)}(t)=\exp\left(-i\omega t n_{x}\sigma_{x}^{(s)}\right)
\exp\left(-i\omega t n_{z}\sigma_{z}^{(s)}\right),\label{Uk}
\end{eqnarray}
for $s=1,2$. Because Eq. (\ref{H0}) is a scalar product, it  is
invariant under rotations and
$U^{\dagger}_{c}(t)H_{0}U_{c}(t)=H_{0}$. This property of the
Heisenberg interaction tremendously simplifies the quantum
operations executed under the protection by continuous dynamical
decoupling. If it were not for this rotational invariance, we
would have to proceed as in Ref. \cite{fanchini022329} and
introduce an auxiliary rotating reference frame, complicating the
procedure. Furthermore, this invariance has another peculiarity:
the same field arrangement that can preserve a quantum memory, can
also, with exactly the same configuration, protect the
quantum-gate operation. No reconfiguration of fields being
necessary during the gate operation is a tremendous
simplification; it certainly improves the prospects for
experimental realization.

To illustrate our protective scheme, we begin by assuming
that the interaction Hamiltonian between the qubit system and the rest of the universe is of the form
\begin{eqnarray}
H_{\rm int}={\bf B}^{(1)}\cdot {\bm \sigma}^{(1)}+{\bf B}^{(2)}\cdot {\bm \sigma}^{(2)},\label{Hint}
\end{eqnarray}
where ${\bf B}^{(s)}=\sum _{m=1} ^{3}B_{m}^{(s)}{\bf
\hat{x}}_{m}$, for $s=1,2$,  with ${\bf \hat{x}}_{1}\equiv{\bf
\hat{x}}$, ${\bf \hat{x}}_{2}\equiv{\bf \hat{y}}$, ${\bf
\hat{x}}_{3}\equiv{\bf \hat{z}}$, and $B_{m}^{(s)}$, for $s=1,2$
and $m=1,2,3$, are Hermitian operators that act on the
environmental Hilbert space. It follows from Eqs. (\ref{Uc}),
(\ref{Uk}), and (\ref{Hint}) that $U_{c}(t)$ satisfies the
requirement for dynamical decoupling \cite{DD}:
\begin{equation}
\int ^{t_{c}}_{0} U^{\dagger}_{c}(t)H_{\rm int}U_{c}(t) dt=0,
\end{equation}
where $t_{c}=2\pi /\omega $. In order to control the intensity of
the exchange interaction, possible candidates for the physical
qubits should be, for example, properly built tunable charge
qubits \cite{schon}. Although, in this particular case, physical
reasoning leads us to assume that each qubit is coupled to its own
environment, we shall, for the sake of completeness, also study
the case of a common environment. For our present purposes, we
assume that the particular form of Eq. (\ref{Hint}) is, in the case of a common environment, given by
\begin{equation}
H_{int}=\left({\bm \sigma}^{(1)} + {\bm \sigma}^{(2)}\right)\cdot
\left({\bm \lambda}B + {\bm\lambda }^\ast
B^\dagger\right),\label{collective}
\end{equation}
where
%\begin{equation}
${\bm B}^{(s)}={\bm \lambda}B+{\bm \lambda}^{\ast}B^{\dagger}$,
%\end{equation}
for $s=1,2$, ${\bm \lambda}$ is an arbitrary complex
three-dimensional  vector, and $B$ is a scalar operator that acts
on the environmental Hilbert space. However, when the qubits are physically located
sufficiently far apart, as for tunable charge
qubits \cite{schon}, it is reasonable to suppose that their
individual surroundings act as uncorrelated, independent
environments. In that case, the particular form we assume for Eq.
(\ref{Hint}) is written as
\begin{eqnarray}
H_{int}&=&{\bm \sigma}^{(1)}\cdot\left({\bm \lambda^{(1)}}B^{(1)} +
{\bm\lambda^{(1)}}^\ast {B^{(1)}}^\dagger\right)\nonumber\\
&+&{\bm \sigma}^{(2)}\cdot\left({\bm \lambda^{(2)}}B^{(2)} +
{\bm\lambda^{(2)}}^\ast {B^{(2)}}^\dagger\right)\label{independent}
\end{eqnarray}
where $B^{(s)}$, for $s=1,2$, acts on the environmental Hilbert
space  of the $s$-th qubit, and ${\bm \lambda^{(s)}}$ is an
arbitrary complex three-dimensional vector for $s=1,2$.

We first consider the point of view of the picture defined by the
unitary transformation $U_{c}(t)$; thus, the total Hamiltonian can
be written
\begin{eqnarray}
H(t)=H_{0}+H_{E}+U^{\dagger}_{c}(t)H_{\rm int}U_{c}(t),\label{Htot}
\end{eqnarray}
where $H_{E}$ is the environmental Hamiltonian and, therefore,
$U^{\dagger}_{c}(t)H_{E}U_{c}(t)=H_{E}$. We represent the
environment of each qubit as a thermal bath of harmonic
oscillators. In the case of a common environment for both qubits,
we consider
%\begin{equation}
$H_{E}=\sum _{k} \omega_{k} {a_{k}}^{\dagger}a_{k}$,
%\end{equation}
where $\omega _{k}$ is the frequency of the $k$-th normal mode  of  the common environment,
and $a_{k}$ and ${a_{k}}^\dagger$ are the annihilation and creation operators, respectively.
In the case of two independent and identical environments, instead of %Eq. (\ref{bath}), we take
above we take $H_{E}=\sum_{s=1}^{2}\sum _{k} \omega_{k}
{a_{k}^{(s)}}^{\dagger}a_{k}^{(s)}$,  where $\omega _{k}$ is the
frequency of the $k$-th normal mode of the $s$-th qubit
environment, and $a_{k}^{(s)}$ and ${a_{k}^{(s)}}^\dagger$ are,
respectively, the annihilation and creation operators. The
frequency $\omega _{k}$ is the same for both independent and
identical environments. Accordingly, we take
%\begin{eqnarray*}
$B_{m}^{(s)}  =  \sum_{k}\left(\lambda_{m}g_{k}^{\ast}a_{k}^{(s)}+
\lambda_{m}^{\ast}g_{k}a_{k}^{(s)\dagger}\right)$,%\end{eqnarray*}
where $g_{k}$ are coupling constants.

Starting from $H(t)$, the Hamiltonian in the ``interaction'' picture is defined as
\begin{eqnarray}
H_{I}(t)=\sum _{s=1} ^{2}\sum _{m=1} ^{3}\sum_{n =1}^{3}R_{m,n}(t)E_{m}^{(s)}(t)
\widetilde{\sigma }_{n}^{(s)}(t).\label{HI}
\end{eqnarray}
where $\sigma_{1}^{(s)}\equiv \sigma_{x}^{(s)}$,
$\sigma_{2}^{(s)}\equiv \sigma_{y}^{(s)}$, $\sigma_{3}^{(s)}\equiv
\sigma_{z}^{(s)}$, $\widetilde{\sigma }_{n}^{(s)}(t)=
U^{\dagger}_{0} (t)\sigma _{n}^{(s)}U_{0} (t)$, for $s=1,2$ and
$n=1,2,3$, with $U_{0}(t)=\exp(-iH_{0}t)$. We have used Eq.
(\ref{Hint}) and defined the operators
$E_{m}^{(s)}(t)=U^{\dagger}_{E}(t)B_{m}^{(s)}U_{E}(t)$, for
$s=1,2$ and $m=1,2,3$, with $U_{E}(t)=\exp(-iH_{E}t)$. The
quantities $U^{\dagger}_{c} (t)\sigma_{m}^{(s)}U_{c} (t)=\sum_{n
=1}^{3} R_{m,n}(t) \sigma_{n}^{(s)}$, for $s=1,2$ and $m=1,2,3$,
are rotations of $\sigma_{m}^{(s)}$, whose matrix elements,
$R_{m,n}(t)$, are real functions of time. We proceed as  in Ref.
\cite{fanchini07} and assume that the absolute temperature is the
same in the surroundings of both qubits and these qubits, as well
as their respective environments, are identical. We then write
down the master equation for the two-qubit reduced density matrix,
$\rho_{I}(t)$, in the Born approximation:
\begin{widetext}
\begin{eqnarray}
\frac{d\rho_{I}(t)}{dt} & = & \sum_{s,s^{\prime}=1}^{2}
\sum_{n,n^{\prime}=1}^{3}\int_{0}^{t}dt^{\prime}\,
\left\{\mathcal{D}_{n,n^{\prime}}^{(s,s^{\prime})}(t,t^{\prime})
[\widetilde{\sigma}_{n}^{(s)}(t),\rho_{I}(t)\widetilde{\sigma}_{n^{\prime}}^{(s^{\prime})}(t^{\prime})]
+[\mathcal{D}_{n,n^{\prime}}^{(s,s^{\prime})}(t,t^{\prime})]^{*}
[\widetilde{\sigma}_{n^{\prime}}^{(s^{\prime})}(t^{\prime})\rho_{I}(t),
\widetilde{\sigma}_{n}^{(s)}(t)]\right\},\label{master}
\end{eqnarray}
\end{widetext}
where we have define the coefficients
\begin{eqnarray*}
\mathcal{D}_{n,n^{\prime}}^{(s,s^{\prime})}(t,t^{\prime}) & = &
\sum_{m=1}^{3}
\sum_{m^{\prime}=1}^{3}R_{m,n}(t)R_{m^{\prime},n^{\prime}}(t^{\prime})
C_{m,m^{\prime}}^{(s,s^{\prime})}(t,t^{\prime}),\end{eqnarray*}
for $n,n^{\prime}=1,2,3$ and $s,s^{\prime}=1,2$,
and\begin{eqnarray*}
C_{m,m^{\prime}}^{(s,s^{\prime})}(t,t^{\prime}) & = &
\mathrm{Tr}_{E} \left\{
E_{m}^{(s)}(t)\rho_{E}E_{m^{\prime}}^{(s^{\prime})}(t^{\prime})\right\}
,\end{eqnarray*} for $m,m^{\prime}=1,2,3$ and  $s,s^{\prime}=1,2$.
$C_{m,m^{\prime}}^{(s,s^{\prime})}(t,t^{\prime})$ is the
correlation function between components $m$ and $m^{\prime}$ of
environmental operators calculated at the same qubit position, as
explained in Ref. \cite{fanchini07}. Here, $\mathrm{Tr}_{E}$
denotes the trace over the environmental degrees of freedom. The
operators $\widetilde{\sigma}_{n}^{(s)}(t)$, for $s=1,2$ and
$n=1,2,3$, can be explicitly obtained as the components of the
following vector relations:
\begin{eqnarray}
U^{\dagger}_{0} (t){\bm \sigma }^{(1)}U_{0} (t)=a(t){\bm \sigma }^{(1)}+b(t){\bm \sigma }^{(2)}\nonumber \\
-c(t)({\bm \sigma }^{(1)}\times {\bm \sigma }^{(2)}),
\end{eqnarray}
and
\begin{eqnarray}
U^{\dagger}_{0} (t){\bm \sigma }^{(2)}U_{0} (t)=a(t){\bm \sigma }^{(2)}+b(t){\bm \sigma }^{(1)}\nonumber \\
-c(t)({\bm \sigma }^{(2)}\times {\bm \sigma }^{(1)}),
\end{eqnarray}
where $a(t)=[1+\cos (4Jt)]/2$, $b(t)=[1-\cos (4Jt)]/2$, and
$c(t)=\sin (4Jt)/2$.  The environmental density matrix, $\rho
_{E}$, is taken as the one for a canonical ensemble constituting a
thermal bath, that is,
%\begin{eqnarray}
$\rho _{E}=\frac{1}{Z}\exp(-\beta H_{E})$,%\label{rhoE}
%\end{eqnarray}
where $Z$ is the partition function, $Z={\mathrm
Tr}_{E}\left[\exp(-\beta H_{E})\right]$.  Here, $\beta =1/k_{B}T$,
$k_{B}$ is the Boltzmann constant, and $T$ is the absolute
temperature of the environment.

We can also write the correlation function as\begin{eqnarray*}
C_{m,m^{\prime}}^{(s,s^{\prime})}(t,t^{\prime}) & = & \Gamma^{(s,s^{\prime})}\mathrm{Tr}_{E}
\left\{ E_{m}^{(s)}(t)\rho_{E}E_{m^{\prime}}^{(s)}(t^{\prime})\right\} ,\end{eqnarray*}
where $\Gamma^{(s,s^{\prime})}=1$ for the case of a single, common
environment, in which case the environmental operators $E_{m}^{(s)}(t)$
are independent of $s$, and $\Gamma^{(s,s^{\prime})}=\delta_{s,s^{\prime}}$
for the case of two identical, uncorrelated environments. Since we
have
\begin{eqnarray*}
E_{m}^{(s)}(t) & = & \sum_{k}\left[\lambda_{m}g_{k}^{\ast}a_{k}^{(s)}e^{-i\omega_{k}t}+
\lambda_{m}^{\ast}g_{k}a_{k}^{(s)\dagger}e^{+i\omega_{k}t}\right]\end{eqnarray*}
and, therefore,

\begin{eqnarray*}
\mathrm{Tr}_{E}\!\!\left\{\! E_{m}^{(s)}(t)\rho_{E}E_{m^{\prime}}^{(s)}(t^{\prime})\!\right\}\! &=& \!
\lambda_{m}\lambda_{m^{\prime}}^{\ast}\sum_{k}\left|g_{k}\right|^{2}n_{k} e^{-i\omega_{k}(t-t^{\prime})}\\
&&\hspace{-1.05cm}+\,\lambda_{m}^{\ast}\lambda_{m^{\prime}}\sum_{k}\left|g_{k}\right|^{2}(1+n_{k})
e^{i\omega_{k}(t-t^{\prime})},
\end{eqnarray*}

%\begin{widetext}
%\begin{eqnarray*}
%\mathrm{Tr}_{E}\left\{ E_{m}^{(s)}(t)\rho_{E}E_{m^{\prime}}^{(s)}
%(t^{\prime})\right\}  & = &
%\lambda_{m}\lambda_{m^{\prime}}^{\ast}\sum_{k}\left|g_{k}\right|^{2}n_{k}\exp[-i\omega_{k}(t-t^{\prime})]
%+
%$\lambda_{m}^{\ast}\lambda_{m^{\prime}}\sum_{k}\left|g_{k}\right|^{2}(1+n_{k})\exp[i\omega_{k}(t-t^{\prime})],
%\end{eqnarray*}
%\end{widetext}
where $ n_{k}=1/[\exp(\beta\omega_{k})-1]$, we obtain
\begin{widetext}
\begin{eqnarray*}
\frac{d\rho_{I}(t)}{dt} & = & \sum_{s,s^{\prime}=1}^{2}
\int_{0}^{t}dt^{\prime}\,\mathcal{T}_{1}^{(s,s^{\prime})}(t-t^{\prime})
[\mathcal{R}^{(s)}(t),\rho_{I}(t)[\mathcal{R}^{(s^{\prime})}
(t^{\prime})]^{\dagger}]+\sum_{s,s^{\prime}=1}^{2}\int_{0}^{t}dt^{\prime}\,
\mathcal{T}_{2}^{(s,s^{\prime})}(t-t^{\prime})
[[\mathcal{R}^{(s)}(t)]^{\dagger},\rho_{I}(t)\mathcal{R}^{(s^{\prime})}(t^{\prime})]\\
 & + & \sum_{s,s^{\prime}=1}^{2}\int_{0}^{t}dt^{\prime}\,
 [\mathcal{T}_{1}^{(s,s^{\prime})}(t-t^{\prime})]^{*}
 [\mathcal{R}^{(s^{\prime})}(t^{\prime})\rho_{I}(t),[\mathcal{R}^{(s)}(t)]^{\dagger}]+
 \sum_{s,s^{\prime}=1}^{2}\int_{0}^{t}dt^{\prime}\,
 [\mathcal{T}_{2}^{(s,s^{\prime})}(t-t^{\prime})]^{*}
 [[\mathcal{R}^{(s^{\prime})}(t^{\prime})]^{\dagger}\rho_{I}(t),\mathcal{R}^{(s)}(t)]\end{eqnarray*}
\end{widetext}
where
\begin{eqnarray*}
\mathcal{R}^{(s)}(t) & = & \sum_{m=1}^{3}\sum_{n=1}^{3}\lambda_{m}R_{m,n}(t)\widetilde{\sigma}_{n}^{(s)}(t),
\end{eqnarray*}
\begin{eqnarray*}
\mathcal{T}_{1}^{(s,s^{\prime})}(t) & = & \Gamma^{(s,s^{\prime})}\sum_{k}\left|g_{k}\right|^{2}n_{k}\exp(-i\omega_{k}t),
\end{eqnarray*}
and
\begin{eqnarray*}
\mathcal{T}_{2}^{(s,s^{\prime})}(t) & = & \Gamma^{(s,s^{\prime})}\sum_{k}\left|g_{k}\right|^{2}(1+n_{k})
\exp(i\omega_{k}t).
\end{eqnarray*}

In the limit in which the number of environmental normal modes per
unit frequency becomes infinite, we define a spectral density as $J(\omega)=\sum_{k}|g_{k}|^{2}
\delta(\omega-\omega_{k})$,
with $\omega\in\left[0,\infty\right)$, and interpret the summations
in $\mathcal{T}_{1}^{(s,s^{\prime})}(t)$ and $\mathcal{T}_{2}^{(s,s^{\prime})}(t)$
as integrals over $\omega$:\begin{eqnarray*}
\mathcal{T}_{1}^{(s,s^{\prime})}(t) & = & \Gamma^{(s,s^{\prime})}
\int_{0}^{\infty}d\omega J(\omega)\frac{\exp(-i\omega t)}{\exp(\beta\omega)-1},\end{eqnarray*}
 and \begin{eqnarray*}
\mathcal{T}_{2}^{(s,s^{\prime})}(t) & = & \left[\mathcal{T}_{1}^{(s,s^{\prime})}(t)\right]^{\ast}+
\Gamma^{(s,s^{\prime})}\int_{0}^{\infty}d\omega J(\omega)\exp(i\omega t),\end{eqnarray*}
for $s,s^{\prime}=1,2$. Here we assume an ohmic spectral density
with a cutoff frequency $\omega_{c}$, namely, $J(\omega)=\eta\omega\exp(-\omega/\omega_{c})$,
where $\eta$ is a dimensionless constant.
\begin{figure} %[htbp]
\includegraphics[width=.48\textwidth]{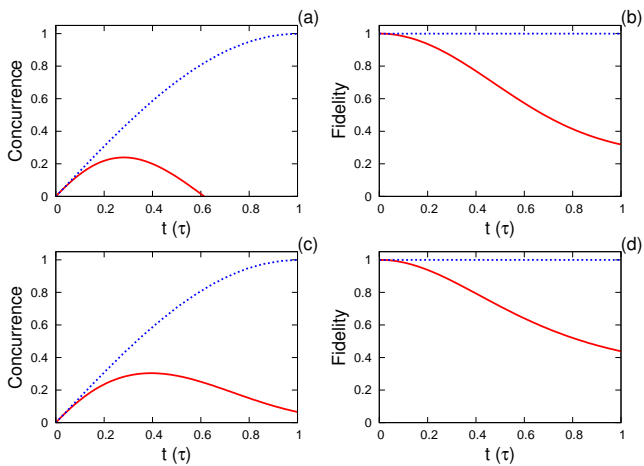}
\caption{(Color Online) \textit{Amplitude Damping plus Dephasing:} In figures (a)
and (b), we show, respectively, the concurrence and the fidelity
for independent environments and, in figures (c) and (d), for the
common environments. The dotted (blue) line and solid (red) line
represent the dynamics of a $\sqrt{\rm{SWAP}}$ quantum gate, with
and without protection, respectively.}\label{fig1}
\end{figure}
To illustrate our method, we consider the protection of an
entangling operation.  We take $J=\pi/8$ in Eq. (\ref {H0}) and
assume that the two qubits are coupled to ohmic environments at
$T=0.2K$, with the coupling constant $\eta=1/20$, and the cut-off
frequency given by $\omega_c\tau=2\pi$, where $\tau=10^{-9}$s. We
consider two uncorrelated classes of errors: amplitude damping and
dephasing. We suppose that
$\rho(0)=|\!\!\uparrow\downarrow\rangle\langle\uparrow\downarrow\!\!|$
and, in Fig. (\ref {fig1}), we show the fidelities and the
concurrences \cite{concurrence} for the protected and unprotected
cases of the $\sqrt{\rm SWAP}$ quantum gate. We consider that the
qubits interact with independent or common environments. For the
protected cases, we take $n_x=28\pi/\tau$ and $n_z=14\pi/\tau$. We
observe, in all protected cases, higher fidelities and
concurrences, as compared to the unprotected cases. In all
examples shown in Fig. (\ref{fig1}), the final fidelities and
concurrences of the protected dynamics are near unity. In fact, in
the protected cases shown, they are greater than $0.998$ and
higher values can be obtained for greater values of $n_x$ and
$n_z$.

To summarize, we present a simplified method to protect a
$\sqrt{{\rm SWAP}}$ quantum  gate from general classes of errors.
Our scheme protects the logical operation at the same time as it
is applied and, using the same control-field arrangement, can
protect a quantum memory or a quantum gate. The flexibility of
using the same control field, in the static and dynamic
situations, greatly improves the prospects for an experimental
realization. Furthermore, since the quantum gates derived from the
exchange interaction alone are universal per se, our methodology
provides the possibility of a totally-protected universal quantum
computation, using continuous dynamical decoupling.

This work has been partly supported by ``Fundação de Amparo à
Pesquisa do Estado de São Paulo (FAPESP)'', Brazil, project number
05/04105-5, and by FAPESP and ``Conselho Nacional de
Desenvolvimento Científico e Tecnológico (CNPq)'', Brazil, through
the ``Instituto Nacional de Ci{\^e}ncia e Tecnologia em
Informa\c{c}{\~a}o Qu{\^a}ntica (INCT-IQ)''.

\end{document}